\documentclass[preprint, authoryear, 12pt]{elsarticle}

 \usepackage{graphicx}

\usepackage{amssymb}
\usepackage{amsmath}
\usepackage{bm}

\usepackage{soul}
\usepackage{color}
\sethlcolor{yellow}

\usepackage[colorlinks=true]{hyperref}

\renewcommand{\hl}[1]{#1}

\journal{Advances in Space Research}

\begin{document}

\begin{frontmatter}

\title{\hl{Optimization of Fault-Tolerant Thruster Configurations for Satellite Control}}

\author[label1]{Yasuhiro Yoshimura\corref{cor}}
\address[label1]{Tokyo Metropolitan University, 6-6 Asahigaoka, Hino, Tokyo 191-0065, Japan}
\cortext[cor]{Corresponding author}
\ead{yyoshi@tmu.ac.jp}

\author[label1]{Hirohisa Kojima}
\ead{hkojima@tmu.ac.jp}

\begin{abstract}
\hl{The fault tolerance of spacecraft actuators significantly affects the reliability of satellites and the likelihood of successful missions. To enhance the fault tolerance of the actuators, this study derives optimal fault-tolerant configurations of fixed thrusters that maximize the controllability of a fully-actuated or underactuated satellite. The proposed method optimizes thrust and torque directions generated by the thrusters. Thus a cost function in terms of the thruster locations and directions is defined as the summation of the generated control forces and torques with respect to the body-fixed frame. The optimal configuration is obtained by the successive use of an energy potential method that is motivated by Thomson's problem. Some numerical examples are provided that show the effectiveness of the proposed formulation and optimization method.}
\end{abstract}


\end{frontmatter}


\section{Introduction}
\label{intro}

The fault tolerance of spacecraft actuators significantly affects satellite reliability and the chances of mission success. One way of enhancing the fault tolerance against actuator failures is to use underactuated control, which enables the driving of satellites into desired states with lower number of inputs than the number of state variables. Underactuated control \hl{of} satellites has been intensively studied, and many control techniques have been proposed~\citep{Krishnan:1994tw,Tsiotras1995,Morin1997,Yoshimura2011,Horri2012}.

In practical situations, however, one of the difficulties in applying underactuated controllers is that input directions are restricted when some actuators fail. When actuators malfunction, the remaining actuators are not necessarily able to generate control torques and/or translational forces in the ideal directions, such as along the principal axes of inertia. Such restriction on the possible input directions makes underactuated controllers somewhat impractical. In other words, underactuated control can enhance the fault tolerance of a satellite as long as the controllability of the satellite is still sufficient even after some actuators have failed. In this context, this study derives the optimal fault-tolerant configurations of \hl{fixed} thrusters that maximize the controllability of underactuated satellites.

A thruster generates a translational force and coupled torque in a single direction due to thruster mechanisms. Although the position and attitude of a satellite can be simultaneously controlled with thrusters, the unilateral control inputs complicate the proof of the controllability of the system. In fact, controllability theorems in the case of restricted input directions have not yet been proposed. \hl{The derivation of the controllability conditions is outside of the scope of this study. Instead, two controllability conditions used in}~\citet{Pena:2000td, MESA} \hl{are applied}.

It is known that four thrusters are necessary for \hl{the attitude control of a fully-actuated satellite} as shown in~\citet{Yoshimura2011,Sidi}. Underactuated controllers, however, enable attitude control with only three thrusters~\citep{MESA}. Although several previous studies have discussed the minimum necessary number of thrusters and their configuration, few studies have considered underactuated control. ~\citet{Pena:2000td} \hl{showed} an optimal and robust 6-thruster configuration that is robust against the failure of a single thruster. ~\citet{Jin1995} also \hl{provided} an optimal thruster configuration that maximizes the margin of safety of the thrusters, with the thruster configuration being designed to attenuate external disturbances sufficiently. For gimbal thrusters, thruster configurations that consume less fuel are studied by~\cite{Saberi2015}. These previous studies focus on disturbance rejection and fuel consumption, and the controllability associated with underactuated control systems is not considered.

The optimization of \hl{fixed-}thruster configurations requires finding optimal thruster locations and directions with respect to a satellite body-fixed frame. \hl{We consider general thruster locations and directions, i.e., the thrusters generate translational forces and their coupled torques.} A cost function for the controllability is built on the condition that translational forces and rotational torques must be able to be generated in any directions. This condition defines the cost function as the sum of the generated control forces and torques with respect to the body-fixed frame. The optimal thruster configuration can be derived using a solution to Thomson's problem, as applied in~\cite{Yoshimura2015}. In this optimization method, by considering the geometric locations and directions of the thrusters as point charges, an arbitrary number of thrusters can be configured at equal distances, maximizing the available control forces and torques in all directions. That is, the controllability of the position and attitude of the satellite after actuator failures is maximized. Furthermore, applying different weights to the point charges can account for underactuated controllability. Some numerical examples are provided to demonstrate the effectiveness of the proposed formulation and the optimization method.

\hl{There a few papers that deal with translational and rotational motion control of an underactuated satellite. Yoshimura et al. showed an open-loop controller that can drive a satellite to arbitrary position and attitude with only four thrusters. Although this method is based on open loop controller, the analytical solution of the translational and rotational motion provides clues for trajectory design of the underactuated satellite. Pong also presented the translational and rotational motion control of an underactuated satellite. In this thesis, common control techniques such as model predictive control are studied for implementing the control of the underactuated satellite.}

The rest of this paper is organized as follows. Section 2 formulates optimal thruster configurations in terms of the controllability of satellite position and attitude. The cost function to be minimized is interpreted as the summation of the inner products of translational forces and control torques. Section 3 demonstrates the optimization of the thruster configurations using a method based on the energy potential method. Section 4 gives some numerical examples of the optimal thruster configurations. Some conclusions are presented in Section 5.

\section{Problem Formulation}

\subsection{Satellite and thruster model}
Thrusters generate translational forces with coupled torques. This paper considers the controllability of a satellite position and attitude using thrusters. The satellite position is considered to be a free-floating state, and orbital motion is not considered. The equations of motion of a free-floating satellite can be expressed with an affine system as:
\begin{eqnarray}
\dot{\bm{x}} = \bm{f}(\bm{x}) + G(\bm{x})\bm{u} \label{eq:affineSystem}
\end{eqnarray}
where
\begin{eqnarray}
\bm{x}&=&\left[ {\begin{array}{*{20}{c}}
x&y&z
&\bm{q}^{T}
&{\dot x}&{\dot y}&{\dot z}
&\bm{\omega}^{T}
\end{array}} \right] \nonumber\label{eq:stateVector}\\
\end{eqnarray}
\begin{eqnarray}
\bm{f}(\bm{x}) = \left[ \begin{array}{c}
\dot{x}\\
\dot{y}\\
\dot{z}\\
0.5 [\tilde{\bm{\omega}}] \bm{q}\\
\bm{0}_{3\times 1}\\
-I^{-1}\left( \bm{\omega} \times I \bm{\omega}\right)
\end{array}\right]\label{eq:fx}
\end{eqnarray}

\begin{eqnarray}
G(\bm{x}) &=& \left[ \begin{array}{cc}
\bm{0}_{3\times 3}&\bm{0}_{3\times 3}\\
\bm{0}_{4\times 3}& \bm{0}_{4\times 3}\\
\frac{1}{m}R_{b/i}^{T}& \bm{0}_{3\times 3}\\
\bm{0}_{3\times 3} & I^{-1}
\end{array}\right]\label{eq:gx}
\end{eqnarray}

\begin{eqnarray}
\bm{u} = \left[ {\begin{array}{*{20}{c}}
\bm{F}_{b}^{T} & \bm{T}_{b}^{T}
\end{array}} \right]^{T} \nonumber \label{eq:inputVector}\\
\end{eqnarray}

The satellite position with respect to an inertial frame is expressed in terms of $x$, $y$ and $z$. The attitude angle is formulated with a quaternion $\bm{q} = \left[ \begin{array}{cccc}q_{1} &q_{2} &q_{3}&q_{4} \end{array}\right]^{T}$, in which $q_{4}$ is the scalar part and the other variables form the vector component. The angular velocity is $\bm{\omega}=\left[ \begin{array}{ccc}
\omega_{x} &\omega_{y} &\omega_{z}      \end{array}\right]^{T}$. The satellite mass is $m$ and the matrix $I$ represents the moment of inertia of the satellite. The matrix $R_{b/i}$ is a directional cosine matrix from the inertial frame to the body-fixed frame. In Eq.~\eqref{eq:fx}, $[\tilde{\bm{\omega}}]$ is defined for kinematics with quaternions as follows.

\begin{eqnarray}
\left[ \tilde{ \bm{\omega}}  \right] = \left[ {\begin{array}{*{20}{c}}
  0&{{\omega _z}}&{ - {\omega _y}}&{{\omega _x}} \\ 
  { - {\omega _z}}&0&{{\omega _x}}&{{\omega _y}} \\ 
  {{\omega _y}}&{ - {\omega _x}}&0&{{\omega _z}} \\ 
  { - {\omega _x}}&{ - {\omega _y}}&{ - {\omega _z}}&0 
\end{array}} \right]
\end{eqnarray}

If $N$ thrusters are fixed to the satellite body and each generates translational force in a single direction, the translational force $ \bm{F}_{b}$ 
and torque $ \bm{T}_{b}$ with respect to the body-fixed frame can be written as 
\begin{eqnarray}
\bm{F}_{b}&=&
  \left[ \begin{array}{*{20}{c}}
  {{F_x}} \\ 
  {{F_y}} \\ 
  {{F_z}} 
\end{array} \right] = \left[ {\begin{array}{*{20}{c}}
  {{\bm{d}_1}}& \cdots &{{\bm{d}_N}} 
\end{array}} \right] \left[ {\begin{array}{*{20}{c}}
  {{f_1}} \\ 
   \vdots  \\ 
  {{f_N}} 
\end{array}} \right] \\
\bm{T}_{b}&=&
  \left[ {\begin{array}{*{20}{c}}
  {{T_x}} \\ 
  {{T_y}} \\    
  {{T_z}} 
\end{array}} \right] = \left[ {\begin{array}{*{20}{c}}
  {{\bm{r}_1} \times {\bm{d}_1}}& \cdots &{{\bm{r}_N} \times {\bm{d}_N}} 
\end{array}} \right] \left[ {\begin{array}{*{20}{c}}
  {{f_1}} \\ 
   \vdots  \\ 
  {{f_N}} 
\end{array}} \right]
\end{eqnarray}
The locations and orientations of each thruster are represented by the position vectors $\bm{r}_{i}$ and thrust directional vectors $\bm{d}_{i}$ ($i=1,\dots,N$), respectively. Note that the thrust magnitudes are constrained to be positive values or zero, i.e., $f_{i}\geq 0$ ($i=1,\dots,N$), due to thruster mechanisms.

\subsection{Controllability}
The controllability of affine systems with unidirectional inputs is studied by~\citet{Goodwine1996}. The theorem provides that a system with positive directional inputs is controllable if the combinations of the positive directional inputs can generate inputs in negative directions. Thus, the resulting system is reduced to a system that can generate control inputs in any direction. That is, underactuated control cannot be explicitly considered.~\citet{Pena:2000td} also provide thruster configuration conditions that can generate control forces and torques in all directions. In these studies, the controllability of a satellite position and attitude is assumed if thrusters can generate translational forces and rotational torques in any direction. Thus, the following two controllability conditions are applied in this study:
\begin{enumerate}
    \item The satellite position and attitude are controllable when translational forces and rotational torques can be generated in any direction.
    \item The satellite position and attitude are controllable when a thruster configuration satisfies the conditions described in~\cite{MESA}.
\end{enumerate}

We refer to \hl{the first} condition as linearization controllability in this study, because satisfying this condition enables feedback linearization. For the sake of simplicity, we separate the state variables as
$\bm{x}_{1}= \left[ {\begin{array}{*{20}{c}}  {x}&{y}&{z}&{\bm{q}^{T}} \end{array}} \right]^{T}$ and $\bm{x}_{2}= \left[ {\begin{array}{*{20}{c}}  {\dot{x}}&{\dot{y}}&{\dot{z}}&{\bm{\omega}^{T}} \end{array}} \right]^{T}$. The system in Eqs.~\eqref{eq:affineSystem}--\eqref{eq:inputVector} is rewritten as
\begin{eqnarray}
\dot{\bm{x}}_{1}&=&\bm{f}_{1}\left(\bm{x}_{1},\bm{x}_{2}\right) \\
\dot{\bm{x}}_{2}&=&\bm{f}_{2}\left(\bm{x}_{2}\right) + G_{2}\left({\bm{x}_{1}}\right)\bm{u}
\end{eqnarray}
where
\begin{eqnarray}
\bm{f}_{1}\left(\bm{x}_{1},\bm{x}_{2}\right) = \left[ \begin{array}{c}
\dot{x}\\
\dot{y}\\
\dot{z}\\
0.5 [\tilde{\bm{\omega}}] \bm{q}
\end{array} \right]
\end{eqnarray}
\begin{eqnarray}
\bm{f}_{2}\left(\bm{x}_{2}\right) = \left[ \begin{array}{c}
\bm{0}_{3\times 1}\\
-I^{-1}\left( \bm{\omega} \times I \bm{\omega}\right)\end{array} \right]
\end{eqnarray}
\begin{eqnarray}
G_{2}\left(\bm{x}_{1}\right) = \left[ \begin{array}{cc}
\frac{1}{m}R_{b/i}^{T}& \bm{0}_{3\times 3}\\
\bm{0}_{3\times 3} & I^{-1}
\end{array}\right]
\end{eqnarray}
Applying the control inputs $\bm{u}=G_{2}^{-1}\left(-\bm{f}_{2}+\bm{\nu}\right)$ cancels the nonlinear term $\bm{f}_{2}$ in the dynamics of the satellite, resulting in
\begin{eqnarray}
\dot{\bm{x}}_{2} = \bm{\nu}
\end{eqnarray}
where $\bm{\nu}$ are the new control inputs after the feedback linearization. Thus, the position and attitude of the satellite can be easily controlled with the feedback linearization. This controllability condition, however, does not consider underactuated control. 

The second condition, which we call underactuated controllability, is described in~\cite{MESA} and explicitly considers underactuated control, with the geometrical condition of a three-thruster configuration and the underactuated attitude control law being derived.
In the reference~\citep{MESA}, it is shown that three thrusters can control the attitude of a free-floating satellite, and this result has been extended to the position and attitude control of a satellite. The controllability of a satellite with three thrusters has not been rigorously proven, but has been verified by the derivation of control methods using three thrusters. It is noted that such a three-thruster configuration can control the satellite attitude, even though the thruster configuration does not satisfy the Goodwine theorem~\citep{Goodwine1996}.

\subsection{Cost function}
This study first obtains optimal thruster configurations based on the linearization controllability condition, and considers the underactuated controllability by applying different weights in the optimization procedure. The linearization controllability condition requires that translational forces and rotational torques can be generated in any direction, and is obtained by applying Goodwine theorem~\citep{Goodwine1996} to the position and attitude control of a satellite with thrusters as described in Eqs.~\eqref{eq:affineSystem}--~\eqref{eq:inputVector}.
\hl{The condition that translational forces and rotational torques can be generated in all directions means that directional vectors of both the translational forces and rotational torques should be equally spanned with respect to the satellite body-fixed coordinate.} In other words, this condition can be \hl{geometrically} expressed in terms of the inner products of force vectors and torque vectors with respect to the body-fixed frame. That is, all combinations of the inner products among the force/torque vectors should be minimized for the optimal thruster configuration, which indicates that force/torque vectors are distributed at equal distance in the body-fixed frame.
The cost function to be minimized is thus defined as
\begin{eqnarray}
J= \sum_{i=1}^{N}\sum_{j=1}^{N}\left( \bm{d}_{i}^{T} \bm{d}_{j}\right)^{2} + \sum_{i=1}^{N}\sum_{j=1}^{N}\left( \bm{T}_{i}^{T} \bm{T}_{j}\right)^{2} \label{eq:costFunction}
\end{eqnarray}
where $\bm{T}_{i}=\bm{r}_{i}\times \bm{d}_{i}$. \hl{In Eq.}~\eqref{eq:costFunction}\hl{, the combination when $i=j$ should be ignored.}
The first term on the right-hand side represents the maximization of the force vectors, and the second term indicates the maximization of the torque vectors.
In other words, this cost function imposes the geometric condition that all pairs of force/torque vectors should be as orthogonal as possible. \hl{Minimizing the cost function $J$ indicates that possible directions that translational forces and rotational torques can be generated are maximized. That is, the possibility for satisfying the linearization controllability condition is maximized even if some of thrusters malfunction.}

\section{Optimal Thruster Configurations}
The optimization procedure for distributing force vectors and torque vectors at equal distance can be dealt with in a similar manner to Thomson's problem~\citep{Thomson:1904gt}, which seeks to find particle locations equally distributed on a unit sphere. One of the techniques \hl{that is} used to solve Thomson's problem is an energy potential method for the point charges, and this method can also be used to optimize configurations of control moment gyros~\citep{Yoshimura2015}.
Although this optimization is performed only for gimbal axes of control moment gyros in~\citet{Yoshimura2015}, the thruster optimization requires both force vectors and torque vectors. Nevertheless, this study shows that the thruster configuration can be optimized by successively using the method in~\citet{Yoshimura2015} with constraints.

\subsection{Energy potential method}
\begin{figure}[tb]
\centering
\includegraphics[width=0.6\textwidth]{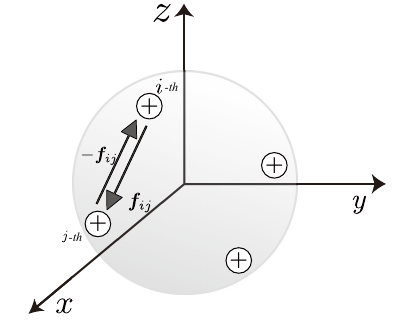}
\caption{\hl{Point charges on a unit sphere}}
\label{fig:thomson}
\end{figure}

To minimize the cost function in Eq.~\eqref{eq:costFunction}, \hl{energy potential method is used in this paper. Although other optimization methods such as genetic algorithms may also be applicable, the energy potential method can be implemented with much less computational burden and iterations than genetic algorithms.} In the energy potential method, considering the endpoints of the force vectors and torque vectors as point charges leads to the same formulation as Thomson's problem.
The energy potential method considers the distribution of the point charges on a unit sphere. \hl{The schematic illustration of the energy potential method is shown in Fig.}~\ref{fig:thomson}. These charges interact with one another and move on the sphere depending on these interacting forces, resulting in a minimum-energy potential state. The magnitudes of the point charges behave as weighting parameters in the optimization process. For example, if equal magnitudes are set, the point charges are equally distributed on the unit sphere surface. 

In this study, the following potential energy formulation for point charges is used.
\begin{eqnarray}
\phi \left( {{\bm{r}_i}} \right) = \frac{A_{i}}{{\left| {{\bm{r}_i}} \right|}}
\end{eqnarray}
where $i=1,\dots,N$ and $A_{i}$ \hl{is the magnitude of a point charge}.
This potential energy represents the Coulomb potential of the interaction, and the following force acts between the $i$-th and the $j$-th charges:
\begin{eqnarray}
{\bm{f}_{ij}} = {A_i}{A_j}\frac{{{\bm{r}_{ij}}}}{{{{\left| {{\bm{r}_{ij}}} \right|}^2}}}  - c_i \dot{\bm{r}}_i \label{eq:f_ij}
\end{eqnarray}
where $\bm{r}_{ij}=\bm{r}_{i}-\bm{r}_{j}$. The second term of the right-hand side of Eq.~\eqref{eq:f_ij} is a virtual damping effect with coefficient $c_{i}$. \hl{This damping effect is introduced so that the point charges converge to a stable state.} Furthermore, the force is projected to confine the particles to the surface:
\begin{eqnarray}
\bm{F}_{ij} = -\left( \bm{f}_{ij} \times \bm{r}_i \right) \times \bm{r}_i\label{eq:interactingF}
\end{eqnarray}

The force $\bm{F}_{ij}$ is calculated for all combinations of the point charges, which then move depending on the forces within a small time step $\Delta t$.
These steps are repeated until the displacements of the point charges become sufficiently small.

\subsection{Optimization of thruster configurations}
The optimization of the thruster configurations is obtained by the successive use of the energy potential method. The force directional vectors, $\bm{d}_{i}$ ($i=1,\dots,N$), are first optimized without constraints. After that, they are fixed, and the torque directional vectors, $\bm{T}_{i}$ ($i=1,\dots,N$), are optimized with the energy potential method under certain constraints. One of the advantages of this energy potential method is that the unilateral constraints on the force directions , i.e., $f_{i}\geq 0 $, can be easily dealt with.

\begin{figure}[tb]
\centering
\includegraphics[width=0.6\textwidth]{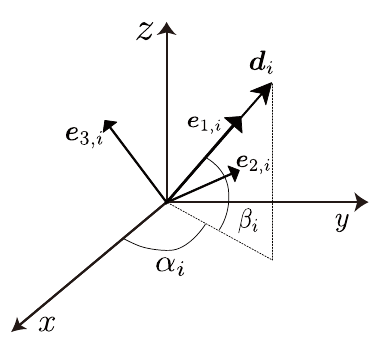}
\caption{\hl{The azimuth and elevation angles of a thruster}}
\label{fig:thr-coordinate}
\end{figure}

Because $\bm{T}_{i}=\bm{r}_{i}\times f_{i}\bm{d}_{i}$, the torque vectors must lie on the orthogonal plane perpendicular to each force directional vector. Furthermore, the position vectors of the thrusters should also be in the same plane so that the generated torques are maximized. Representing the force directional vectors with an azimuth angle, $\alpha_{i}$, and an elevation angle, $\beta_{i}$, as $\bm{d}_{i}= {\left[ {\begin{array}{*{20}{c}}
  {\cos{\beta}_{i} \cos{\alpha}_{i}}&{\cos{\beta}_{i} \sin{\alpha}_{i} }&{\sin{\beta}_{i}} 
\end{array}} \right]^T}$ \hl{as shown in Fig.}~\ref{fig:thr-coordinate}, we configure the basis vectors $\{ \bm{e}_{1,i}, \bm{e}_{2,i}, \bm{e}_{3,i}\}$ as follows:
\begin{eqnarray}
\bm{e}_{1,i} &=& \bm{d}_{i}\\
\bm{e}_{2,i} &=& {\left[ {\begin{array}{*{20}{c}}
  {-\sin{\alpha}_{i}}&{\cos{\alpha}_{i} }&{0} 
\end{array}} \right]^T} \\
\bm{e}_{3,i} &=& \bm{e}_{1,i} \times \bm{e}_{2,i}
\end{eqnarray}

The torque generated by the $i$-th thruster is rewritten in terms of the basis vectors as
\begin{eqnarray}
\bm{T}_{i} = f_{i}\left[ {\begin{array}{*{20}{c}}
  {{\bm{e}_{2,i}}}&{{\bm{e}_{3,i}}} 
\end{array}} \right]\left[ {\begin{array}{*{20}{c}}
  {\cos \gamma_{i} } \\ 
  {\sin \gamma_{i} } 
\end{array}} \right]
\end{eqnarray}
where $\gamma_{i}$ is the phase of the torque vector in the $\bm{e}_{2,i}$--$\bm{e}_{3,i}$ plane. 
Because of orthogonality constraints, the optimization procedure of the torque vectors requires projecting the force onto the $\bm{e}_{2,i}$--$\bm{e}_{3,i}$ plane.
The projected force in Eq.~\eqref{eq:interactingF} is further restricted within the $\bm{e}_{2,i}$--$\bm{e}_{3,i}$ plane as
\begin{eqnarray}
\bm{F}'_{ij} = -\left( \bm{F}_{ij} \times \bm{e}_{1,i} \right) \times \bm{e}_{1,i}\label{eq:Fprime}
\end{eqnarray}
This interacting force includes the orthogonality constraints and is used for the optimization of the torque vectors.

The total torques generated by the $N$ thrusters are then described as follows:
\begin{eqnarray}
\bm{T} =\sum_{i=1}^{N}\bm{T}_{i}\\
= E\left[ {\begin{array}{*{20}{c}}
  {\cos {\gamma _1}} \\ 
  {\sin {\gamma _1}} \\ 
   \vdots  \\ 
  {\sin {\gamma _N}} 
\end{array}} \right]
\end{eqnarray}
\hl{where} $E=\left[ {\begin{array}{*{20}{c}}
  {{f_1}{\bm{e}_{2,1}}}&{{f_1}{\bm{e}_{3,1}}}& f_{2}\bm{e}_{2,2}&f_{2}\bm{e}_{3,2}&\cdots &f_{N}\bm{e}_{2,N}&{{f_N}{\bm{e}_{3,N}}} 
\end{array}} \right]$.
The possible torque directions generated by the $N$ thrusters can be quantitatively evaluated with the following function
\begin{eqnarray}
J_{T}=\det{\left(EE^{T}\right)}
\end{eqnarray}
Since the matrix $E$ includes only azimuth and elevation angles of the thrust directional vectors, the optimization of them determines the 
upper bound of the torque optimization.

\begin{figure}[tb]
\centering
\includegraphics[width=1.0\textwidth]{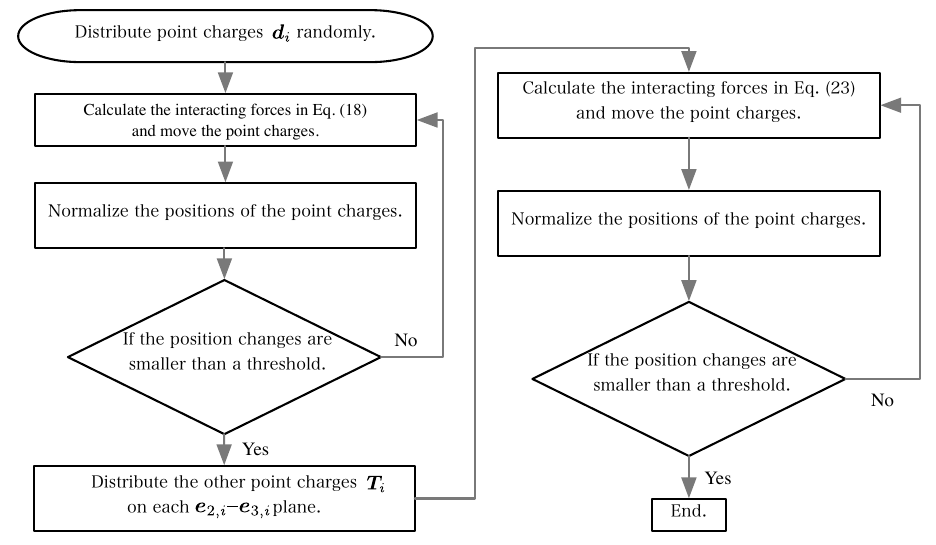}
\caption{The optimization procedure of a thruster configuration}
\label{fig:flow}
\end{figure}

The optimization of an $N$-thruster configuration can be achieved by the procedure in Fig.~\ref{fig:flow}.
This optimization method utilizes the solution to Thomson's problem, and shows that an extension of the method in~\cite{Yoshimura2015} can be used to obtain the optimal thruster configurations. \hl{It should be noted that the proposed method optimizes the both directions of translational forces and rotational torques. Thus the thruster locations are not uniquely determined, i.e., there remains the degrees of freedom for the thruster locations. This is not disadvantage, but the advantage of the proposed method, because the degree of freedom for the thruster locations can consider other criteria such as preventing contamination by the thruster plume.}

The preceding discussion is based on the optimization of the linearization controllability, which is maximized when translational forces and rotational forces are equally oriented in all directions.
On the other hand, the thruster configuration condition discussed in~\cite{MESA} requires control torques in arbitrary directions of the $x$--$y$ plane and unilateral torque along the $z$ axis of the satellite body-fixed frame. The maximization of this condition is realized \hl{by introducing external point charges} in the optimization procedure. For instance, weighting on positive torque along $z$ axis is performed by setting a fixed external point charge that interacts only with the point charges of torques. The external point charge functions to move the other point charges of the torque vectors to the positive $z$ direction. This indicates that the point charges of the thrusters are distributed to generate the positive control torque along the $z$ axis. Thus the underactuated controllability condition can be considered by applying weightings in the optimization procedure.

\section{Numerical Examples}
\subsection{{Optimal thruster configuration for linearization controllability condition}}
This section shows some examples of optimal thruster configurations to verify the proposed optimization method. \hl{In practical operations of satellites, six or more thrusters are necessary}~\cite{Sidi,Tsuda2013,Patton2010}. \hl{Since the position and attitude control of satellites is considered in this paper, the first example relates to the optimization of a 12-thruster configuration whose number is the same as}~\cite{Tsuda2013}.The magnitudes of the charges and the damping coefficients are set to 50.0 and 3.0, respectively. \hl{It is noted that these coefficients are heuristically determined, because the magnitudes of the point charges do not have physical meaning in the thruster configuration.}

\begin{figure}[tb]
\centering
\includegraphics{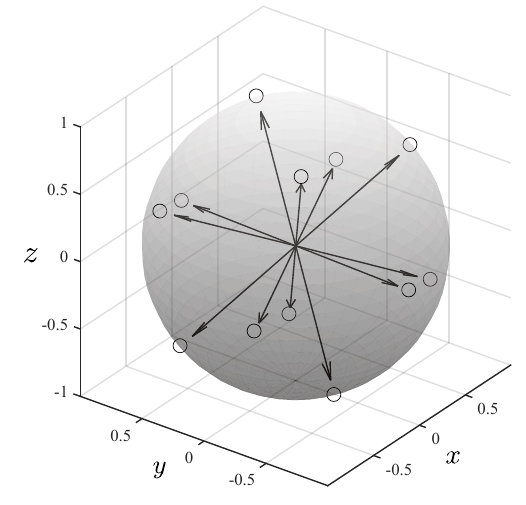}
\caption{\hl{Optimal configuration of thrust directional vectors considering the linearization controllability}}
\label{fig:case1_f}
\end{figure}

\begin{figure}[tb]
\centering
\includegraphics{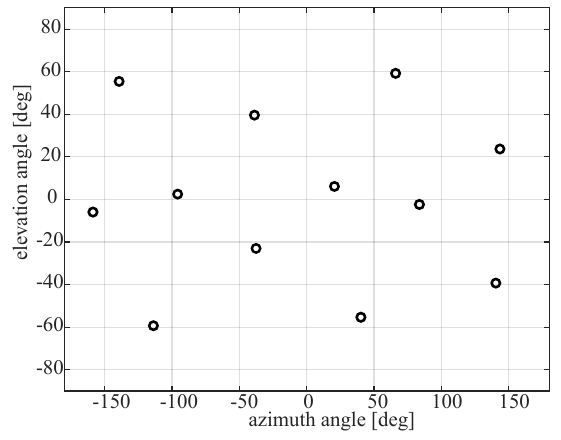}
\caption{\hl{Optimal azimuth and elevation angles of thrust directional vectors considering the linearization controllability condition}}
\label{fig:case1_aziEle_f}
\end{figure}

\begin{figure}[tb]
\centering
\includegraphics{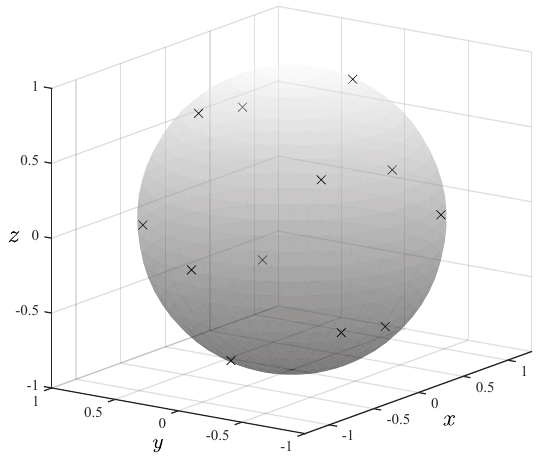}
\caption{\hl{Optimal configuration of thrust directional torque vectors considering the linearization controllability}}
\label{fig:case1_T}
\end{figure}

Figure~\ref{fig:case1_f} shows the position of the distributed point charges, with the circle symbols representing the charges that describe the endpoints of the thrust directional vectors $\bm{d}_{i}$ ($i=1,\dots,12$). Table~\ref{tab:case1_angle_ff} summarizes the angles among the thrust directional vectors, and the minimum angle is 63.4 deg. The lower triangular components are omitted for simplicity. These figure and table show that the distance of all point charges are equally distributed on the sphere. The corresponding azimuth and elevation angles are shown in Fig.~\ref{fig:case1_aziEle_f}. This figure demonstrates that an equal-distance distribution on the sphere is not equivalent to an equal distribution of the azimuth and elevation angles, which is the one of the key difficulties in solving Thomson's problem. The distribution of the torque direction vectors is represented in Fig.~\ref{fig:case1_T}, and these are also successfully distributed so that they can be oriented in any direction.
The cost function results in $J=0.4$, in which both the first term and second term in Eq.~\eqref{eq:costFunction} are 0.4472.
Thus these results show that the thruster configuration maximizes the linearization controllability of the position and attitude of the satellite.

\begin{table}[tb]
\begin{center}
\caption{Angles among the thrust directional vectors}\label{tab:case1_angle_ff}
\begin{tabular}{ccccccccccccc} 
\hline\hline
&$\bm{d}_{1}$ & $\bm{d}_{2}$ &$\bm{d}_{3}$ &$\bm{d}_{4}$ &$\bm{d}_{5}$ &$\bm{d}_{6}$ &$\bm{d}_{7}$ &$\bm{d}_{8}$ &$\bm{d}_{9}$ &$\bm{d}_{10}$ &$\bm{d}_{11}$ &$\bm{d}_{12}$\\ \hline
$\bm{d}_{1}$&   0.0  & 116.6  & 180.0  &  63.4  &  63.4 &  116.6  &  63.4 &  116.6 &  116.6 &  116.6 &   63.4   & 63.4\\
$\bm{d}_{2}$&   &   0.0 &  63.4  & 63.4 & 116.6  &116.6  & 63.4  &116.6 & 63.4   &63.4  &116.6 & 180.0\\
$\bm{d}_{3}$&   & & 0.0  &116.6 & 116.6  & 63.4 & 116.6  & 63.4&  63.4 &  63.4 & 116.6 & 116.6\\
$\bm{d}_{4}$&   &  & &     0.0  &116.6 & 116.6   &63.4 & 180.0  &116.6&   63.4  & 63.4 & 116.6\\
$\bm{d}_{5}$&   &  &   &   &      0.0  &116.6   &63.4 &  63.4  & 63.4 & 180.0  &116.6  & 63.4\\
$\bm{d}_{6}$&    &  & & &  & 0.0  &180.0   &63.4  &116.6 & 63.4  & 63.4   &63.4\\
$\bm{d}_{7}$&    &   &  & & &  &  0.0  &116.6   &63.4 &116.6  &116.6 & 116.6\\
$\bm{d}_{8}$&    & &   && & & &       0.0   &63.4  &116.6 &116.6  & 63.4\\
$\bm{d}_{9}$&   &  &  && & & &   &  0.0  &116.6  &180.0  &116.6\\
$\bm{d}_{10}$&    & & & & & & & & & 0.0  & 63.4  &116.6\\
$\bm{d}_{11}$&    & & & & & & & & & &  0.0  & 63.4\\
$\bm{d}_{12}$&    & & & & & & & & & & &  0.00\\
\hline \hline
\end{tabular}
\end{center}
\end{table}

\subsection{Optimal thruster configuration for underactuated controllability condition}
The next example considers the underactuated controllability condition. As stated above, the condition in~\cite{MESA} requires control torques along the $z$ axis in unilateral direction. The optimization is performed by setting an external point charge at ${\left[ {\begin{array}{*{20}{c}}
  {0.0}&{0.0}&{ - 1.5} 
\end{array}} \right]^T}$. The magnitude of the external point charge is set to 3.0, whereas that of the thrust directional vectors and torque vectors are $A_{i}=50.0$ with the damping coefficients $c_{i}=3.0$ ($i=1,\dots,12$). \hl{It is noted that, as described in Section 3.2, the position and magnitude of the external point charge function as a weighting parameter for the optimization. Thus this external point charge weights the control torques along the $z$ axis.}

\begin{figure}[tb]
\centering
\includegraphics{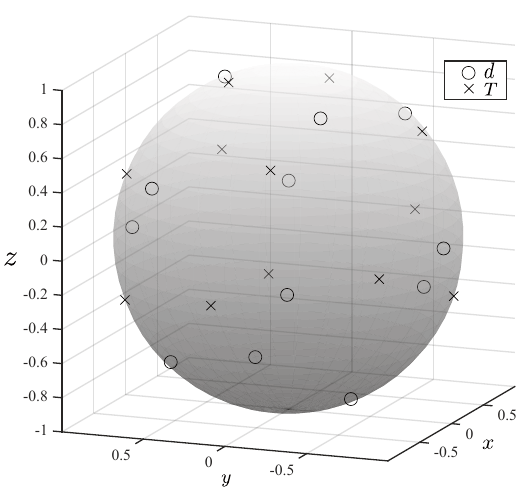}
\caption{Optimal configuration of thrust directional vectors and torque vectors considering the underactuated controllability setting the external point magnitude to 3.0}
\label{fig:case2_fT}
\end{figure}

\begin{figure}[tb]
\centering
\includegraphics{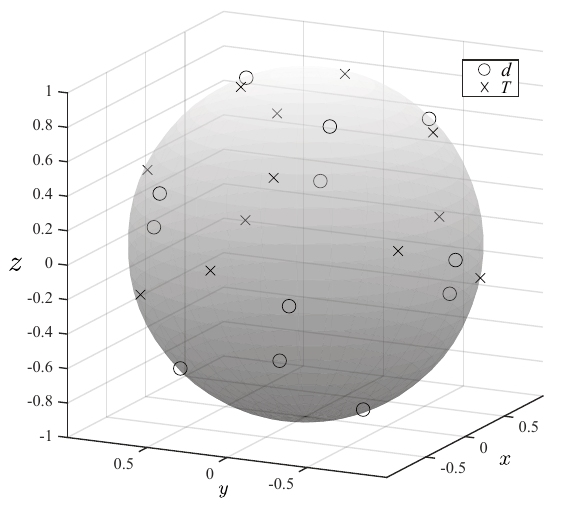}
\caption{Optimal configuration of thrust directional vectors and torque vectors considering the underactuated controllability setting the external point magnitude to 6.0}
\label{fig:case2_fT_b}
\end{figure}

Figure~\ref{fig:case2_fT} shows the optimal configuration of thrust directional vectors and torque vectors considering the underactuated controllability, in which the cycle and \hl{cross} symbols represent the endpoints of the thrust vectors and torque vectors, respectively. The point charges of the torque vectors are not equal distribution, but are allocated along the $z$ axis. Thus the external point charge successfully functions to weight along the positive direction of the $z$ axis.
The cost function becomes $J = 0.4825$, which is larger than the previous example, because the torque vectors are weighted along the $z$ axis.
In fact, the first term of the cost function is 0.4472, whereas the second term is 0.5315. The sum of the $z$ components of the torque vectors also shows the weighted optimization as $\sum_{i=1}^{12}\bm{T}_{z,i}=2.3683$, which verifies the thruster configuration is optimized so that the torque along the $z$ axis can be generated even if a few thrusters fail.
Figure~\ref{fig:case2_fT_b} represents the optimization result setting the magnitude of the external point to 6.0, and the other conditions are the same. Due to the larger magnitude of the external point charge, the torque vectors are further weighted along the positive $z$ direction. The magnitudes of the point charges thus behave as the optimization weights.

\subsection{Fault-tolerance of optimal thruster configurations}
\begin{figure}[tb]
\centering
\includegraphics{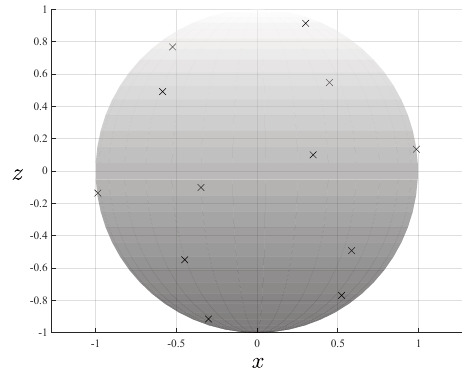}
\caption{Optimal configuration of torque directional vectors considering the linearization controllability}
\label{fig:case1_T_XZ}
\end{figure}

\begin{figure}[tb]
\centering
\includegraphics{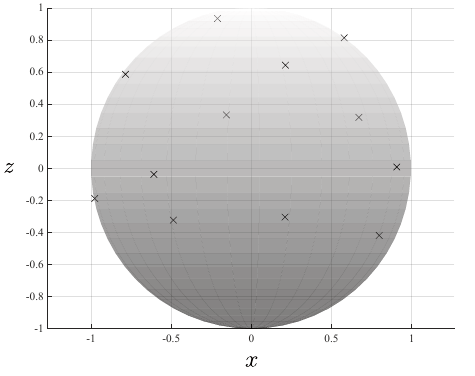}
\caption{Optimal configuration of torque directional vectors considering the underactuated controllability setting the external point magnitude to 3.0}
\label{fig:case2_T_XZ}
\end{figure}

Although both thruster configurations in Section 4.1 and 4.2 consist of 12 thrusters, the first optimization example considers the linearization controllability, whereas the second one considers the underactuated controllability. The optimal configuration considering the underactuated controllability is robust against the malfunctions of the thrusters that generate control torques along $z$ axis. Figures~\ref{fig:case1_T_XZ} and~\ref{fig:case2_T_XZ} show the optimization results of the torque directional vectors in the $x$--$z$ plane. The thruster configuration in Fig.~\ref{fig:case1_T_XZ} has 6 thrusters that generate positive torque along $z$ axis, whereas the thruster configuration in Fig.~\ref{fig:case2_T_XZ} has 7 thrusters. This means that the latter configuration can be controllable in terms of underactuated control even if 6 thrusters of them malfunction, which verify the effectiveness of the proposed method.

\section{Conclusions}
This study presents optimal fault-tolerant configurations of thrusters that maximize the position and attitude controllability of a spacecraft. The controllability accounts for underactuated control with a few thrusters, and the optimal thruster configurations are fault-tolerant even when a thruster malfunction occurs. The optimization of the thruster configuration is reduced to a similar problem to Thomson’s problem. The optimal thruster configuration is then derived by successively using the energy potential method. The proposed method is applicable to both non-underactuated and underactuated systems depending on the weights applied during the optimization procedure. Numerical examples have been provided that verify the effectiveness of the proposed method.






 \bibliographystyle{elsarticle-harv}
\biboptions{authoryear}
\bibliography{my.bib}

\end{document}